\documentclass[a4paper,runningheads]{llncs}

\usepackage[english]{babel}
\usepackage[utf8]{inputenc}
\usepackage[T1]{fontenc}
\usepackage{lmodern}
\usepackage[colorlinks=true]{hyperref}
\usepackage{microtype}
\usepackage{xspace}

\usepackage[bold,small]{complexity}
\usepackage{psystems}

\usepackage{amsthm}
\usepackage{amsmath}
\usepackage{amssymb}

\newcommand{\QSAT}{\textrm{QSAT}\xspace}
\newcommand{\set}[1]{\{#1\}}
\newcommand{\BigQ}{\vec{Q}}
\newcommand{\true}{\mathsf{t}}
\newcommand{\false}{\mathsf{f}}
\newcommand{\yes}{\mathsf{yes}}
\newcommand{\no}{\mathsf{no}}
\newcommand{\rubbish}{\#}
\newcommand{\reset}{\spadesuit}
\newcommand{\pos}{\mathsf{pos}}
\newcommand{\npos}{\mathsf{neg}}
\newcommand{\rset}{\mathsf{end}}

\title{Solving QSAT in Sublinear Depth}

\author{Alberto Leporati\inst{1} \and Luca Manzoni\inst{1} \and Giancarlo Mauri\inst{1} \and\\
  Antonio E. Porreca\inst{1}\inst{2} \and Claudio Zandron\inst{1}}
\authorrunning{A. Leporati \and L. Manzoni \and G. Mauri \and A.E. Porreca \and C. Zandron}
\institute{Dipartimento di Informatica, Sistemistica e Comunicazione\\
  Universit\`a degli Studi di Milano-Bicocca\\
  Viale Sarca 336/14, 20126 Milano, Italy\\
  $\{\email{leporati},\email{luca.manzoni},\email{mauri},
  \email{porreca},\email{zandron}\}$\email{@disco.unimib.it}
  \and
  Aix Marseille Université, Université de Toulon, CNRS, LIS, Marseille, France\\
  \email{antonio.porreca@lis-lab.fr}}

\begin{document}

\maketitle

\begin{abstract}
Among $\PSPACE$-complete problems, \QSAT, or \emph{quantified SAT}, is one of the most used to show that the class of problems solvable in polynomial time by families of a given variant of P~systems includes the whole $\PSPACE$. However, most solutions require a membrane nesting depth that is \emph{linear} with respect to the number of variables of the \QSAT instance under consideration. While a system of a certain depth is needed, since depth $1$ systems only allows to solve problems in $\P^{\boldsymbol\#\P}$, it was until now unclear if a linear depth was, in fact, necessary. Here we use P~systems with active membranes with charges, and we provide a construction that proves that \QSAT can be solved with a \emph{sublinear} nesting depth of order $\frac{n}{\log n}$, where $n$ is the number of variables in the quantified formula given as input.
\end{abstract}

\section{Introduction}
\label{sec:intro}

The solution of the quantified SAT problem (\QSAT) by means of P~Systems with a ``deep'' membrane structure is a common way to prove that uniform families of P~systems are able to solve all the problems in $\PSPACE$~\cite{Alhazov2003a,Sosik2003a}. Most of these approaches exploit the fact that, to verify the validity of a quantified formula $\varphi$, it is sufficient to ``dedicate'' a level of the membrane structure to a single quantifier, with each membrane acting either as an \emph{or} gate (for existential quantifiers) or as an \emph{and} gate (for universal quantifiers). The result is a membrane structure whose depth is, in the case of uniform families, linear with respect to the number of variables in $\varphi$.

We already know that part of the power of P~systems with active membranes with charges is given by the depth of the membrane structure, that allows them to generate finely partitioned structures otherwise impossible to attain with depth $1$ systems. In fact, depth $1$ systems have been proved~\cite{Leporati2014d} to be limited to solve in polynomial time the problems in $\P^{\boldsymbol\#\P}$, a class which is conjecturally smaller than $\PSPACE$. Furthermore, even with constant depth the currently known problems that can be solved all reside inside the counting hierarchy~\cite{Leporati2014e} and we conjecture this inclusion to actually be an upper bound on the computational power of constant depth P~systems. Even in other models of P~systems, like tissue P~systems or depth $1$ P~systems with antimatter, the class $\P^{\boldsymbol\#\P}$ provides a strict upper bound on the computational power~\cite{Leporati2017a,Leporati2017e}.

Since the nesting depth of the membrane structure in the presence of division is such an important factor, it is somewhat surprising that there is a very large gap in our knowledge: the computational power endowed to P~systems by a sublinear (but more than constant) depth is currently unknown. Existing characterisations show that those kinds of systems contain at least the whole counting hierarchy $\CH$ and are limited above by $\PSPACE$. Their exact characterisation is, however, currently unknown. Also in traditional complexity theory the landscape of classes inhabiting the space between $\CH$ and $\PSPACE$ is surprisingly empty: does that mean that everything will collapse to $\CH$ or will reach $\PSPACE$ or that there are new, unknown, complexity classes that are naturally characterised by uniform families of P~systems?

Here we start our investigation by proving that the linear depth (with respect to the number of variables) usually employed to solve \QSAT is actually unnecessary and that, in fact a sublinear depth of $O(\frac{n}{\log n})$ is sufficient. We prove this result by ``compressing'' the usual membrane structure and delegating part of its duties to the internal membranes. Each one of them is, in fact, able to simulate a membrane sub-structure that is traditionally of logarithmic depth. While this first result seems, at a first glance, to suggest that even smaller depths might be sufficient to solve \QSAT, we notice that it is unlikely that this same technique can be employed to further reduce the nesting depth: to obtain further results, different techniques might be needed.

\section{Basic Notions}
\label{sec:basic-notions}

For an introduction to membrane computing and the related notions of formal language theory and multiset processing, we refer the reader to \emph{The Oxford Handbook of Membrane Computing}~\cite{Paun2010a}. Here
we recall the formal definition of P~systems with active membranes using weak non-elementary division rules~\cite{Paun2001a,Zandron2008a}.

\begin{definition}
  A \emph{P~system with active membranes with weak non-elementary division rules} of initial degree~$d \ge 1$ is a tuple
  \begin{align*}
    \Pi = (\Gamma, \Lambda, \mu, w_{h_1}, \ldots, w_{h_d}, R)
  \end{align*}
  where:
  \begin{itemize}
    \item $\Gamma$ is an alphabet, i.e., a finite non-empty set of symbols, usually called \emph{objects};
    \item $\Lambda$ is a finite set of labels;
    \item $\mu$ is a membrane structure (i.e., a rooted \emph{unordered} tree, usually represented by nested brackets) consisting of~$d$ membranes labelled by elements of~$\Lambda$ in a one-to-one way;
    \item $w_{h_1}, \ldots, w_{h_d}$, with~$h_1, \ldots, h_d \in \Lambda$, are multisets (finite sets with multiplicity) of objects in~$\Gamma$, describing the initial contents of each of the~$d$ regions of~$\mu$;
    \item $R$ is a finite set of rules.
  \end{itemize}
\end{definition}

\noindent
Each membrane possesses, besides its label and position in~$\mu$, another attribute called \emph{electrical charge}, which can be either neutral~($0$), positive~($+$) or negative~($-$) and is always neutral before the beginning of the computation.

The rules in~$R$ are of the following types:
\begin{enumerate}
  \item[(a)] \emph{Object evolution rules}, of the form~$\pevolve{h}{\alpha}{a}{w}$. \\
  They can be applied inside a membrane labelled by~$h$, having charge~$\alpha$ and containing an occurrence of the object~$a$; the object~$a$ is rewritten into the multiset~$w$ (i.e.,~$a$ is removed from the multiset in~$h$ and replaced by the objects in~$w$) without changing the charge of $h$.

  \item[(b)] \emph{Send-in communication rules}, of the form $a \, [\;]_h^\alpha \to [b]_h^\beta$. \\
  They can be applied to a membrane labelled by $h$, having charge $\alpha$ and such that the external region contains an occurrence of the object $a$; the object $a$ is sent into $h$ becoming $b$ and, simultaneously, the charge of $h$ is changed to~$\beta$.

  \item[(c)] \emph{Send-out communication rules}, of the form~$\psendout{h}{\alpha}{a}{\beta}{b}$. \\
  They can be applied to a membrane labelled by~$h$, having charge~$\alpha$ and containing an occurrence of the object~$a$; the object~$a$ is sent out from~$h$ to the outside region becoming~$b$ and, simultaneously, the charge of~$h$ becomes~$\beta$.

  \item[(e)] \emph{Elementary division rules}, of the form~$\pdivide{h}{\alpha}{a}{\beta}{b}{\gamma}{c}$ \\
  They can be applied to a membrane labelled by~$h$, having charge~$\alpha$, containing an occurrence of the object~$a$ but having no other membrane inside (an \emph{elementary membrane}); the membrane is divided into two membranes having label~$h$ and charges~$\beta$ and~$\gamma$; the object~$a$ is replaced, respectively, by~$b$ and~$c$, while the other objects of the multiset are replicated in both membranes.

  \item[(f')] \emph{Weak non-elementary division rules, of the form~$\pdivide{h}{\alpha}{a}{\beta}{b}{\gamma}{c}$} \\
  They can be applied to a membrane labelled by~$h$, having charge~$\alpha$, and containing an occurrence of the object~$a$, even if it contains further membranes; the membrane is divided into two membranes having label~$h$ and charges~$\beta$ and~$\gamma$; the object~$a$ is replaced, respectively, by~$b$ and~$c$, while the rest of the contents (including whole membrane substructures) is replicated in both membranes.
\end{enumerate}

A computation step changes the current configuration according to the following set of principles:
\begin{itemize}
  \item Each object and membrane can be subject to at most one rule per  step, except for object evolution rules: inside each membrane, several evolution rules can be applied simultaneously.
  \item The application of rules is \emph{maximally parallel}: each object appearing on the left-hand side of evolution, communication, or division rules must be subject to exactly one of them (unless the current charge of the membrane prohibits it). Analogously, each membrane can only be subject to one communication or division rule (types (b)--(f')) per computation step; these rules will be called \emph{blocking rules} in the rest of the paper. In other words, the only objects and membranes that do not evolve are those associated with no rule, or only to rules that are not applicable due to the electrical charges.
  \item When several conflicting rules can be applied at the same time, a nondeterministic choice is performed; this implies that, in general, multiple possible configurations can be reached after a computation step.
  \item In each computation step, all the chosen rules are applied simultaneously in an atomic way. However, in order to clarify the operational semantics, each computation step is conventionally described as a sequence of micro-steps whereby each membrane evolves only after its internal configuration (including, recursively, the configurations of the membrane substructures it contains) has been updated. For instance, before a membrane division occurs, all chosen object evolution rules must be applied inside it; this way, the objects that are duplicated during the division are already the final ones.
  \item The outermost membrane (the root of the tree) cannot be divided and any object sent out from it cannot re-enter the system again.
\end{itemize}
A \emph{halting computation} of the P~system~$\Pi$ is a finite sequence~$\compC = (\confC_0, \ldots, \confC_k)$ of configurations, where~$\confC_0$ is the initial configuration, every~$\confC_{i+1}$ is reachable from~$\confC_i$ via a single computation step, and no rules of~$\Pi$ are applicable in~$\confC_k$.

P~systems can be used as language \emph{recognisers} by employing two distinguished objects~$\yes$ and~$\no$: we assume that all computations are halting, and that either one copy of object~$\yes$ or one of object~$\no$ is sent out from the outermost membrane, and only in the last computation step, in order to signal acceptance or rejection, respectively. If all computations starting from the same initial configuration are accepting, or all are rejecting, the P~system is said to be \emph{confluent}.

In order to solve decision problems (or, equivalently, decide languages), we use \emph{families} of recogniser P~systems $\familyPi = \{ \Pi_x : x \in \Sigma^\star \}$. Each input~$x$ is associated with a P~system~$\Pi_x$ deciding the membership of~$x$ in a language $L \subseteq \Sigma^\star$ by accepting or rejecting. The mapping~$x \mapsto \Pi_x$ must be efficiently computable for inputs of any length, as discussed in detail in~\cite{Murphy2011a}.
\begin{definition}
  \label{def:uniform}
  A family of P~systems~$\familyPi = \{ \Pi_x : x \in \Sigma^\star \}$ is \emph{(polynomial-time) uniform} if the mapping~$x \mapsto \Pi_x$ can be computed by two polynomial-time deterministic Turing machines~$E$ and~$F$ as follows:
  \begin{itemize}
    \item $F(1^n) = \Pi_n$, where~$n$ is the length of the input~$x$ and~$\Pi_n$ is a common P~system for all inputs of length~$n$ with a distinguished input membrane.
    \item $E(x) = w_x$, where~$w_x$ is a multiset encoding the specific input~$x$.
    \item Finally,~$\Pi_x$ is simply~$\Pi_n$ with~$w_x$ added to a specific membrane, called the  \emph{input membrane}.
  \end{itemize}
  The family~$\familyPi$ is said to be (polynomial-time) semi-uniform if there exists a single deterministic polynomial-time Turing machine~$H$ such that~$H(x) = \Pi_x$ for each~$x \in \Sigma^\star$.
\end{definition}

Any explicit encoding of~$\Pi_x$ is allowed as output of the construction, as long as the number of membranes and objects represented by it does not exceed the length of the whole description, and the rules are listed one by one. This restriction is enforced in order to mimic a (hypothetical) realistic process of construction of the P~systems, where membranes and objects are presumably placed in a constant amount during each construction step, and require actual physical space proportional to their number; see also~\cite{Murphy2011a} for further details on the encoding of P~systems.

\subsection{Polynomial Charges}
\label{sec:toolkit}

As shown in~\cite{Leporati2017b}, it is possible to expand the traditional model of membranes with charges by using a polynomial amount of charges instead of only the usual three. When a polynomial slowdown is acceptable, like in the situation under study, the traditional and the enhanced model have the same computational power with the same membrane nesting depth. Therefore, the construction provided here assumes the presence of a polynomial amount of charges, but the results also hold for the traditional model with three charges. In particular, in this extended model the definition of a P~system is enriched with a finite set $\Psi$ that defines which charges can assume a membrane;  the traditional case, $\Psi = \set{-,0,+}$. Since a polynomial-time uniform family of this kind of P~systems is still constructed using a pair of polynomial-time deterministic Turing machines, also the size of the set of charges is polynomially bounded with respect to the input size. In the rest of the paper for clarity we will represent this extended charges with tuples of various lengths. Nonetheless, each tuple is to be considered as a unique object taken from the set of charges.

\section{Construction}
\label{sec:construction}

Let $\varphi = Q_1Q_2\ldots Q_n \varphi^\star$ be a quantified Boolean formula with $n$ quantifiers over $n$ variables $V = \set{x_1, \ldots, x_n}$ and $\varphi^\star$ its non-quantified version in conjunctive normal form; i.e., $\varphi^\star = C_1 \land C_2 \land \ldots \land C_m$ where each $C_j$, for $1 \le j \le m$, is a disjunctive clause. In particular, we will deal with clauses composed of exactly three literals, in order to obtain a number of possible clauses that is polynomial with respect to the number of variables~\cite{Papadimitriou1993a}. Finally, we denote by $\pos(C_j)$ the set of variables that appear as positive literals in $C_j$ and by $\npos(C_j)$ the set of variables that appears in $C_j$ as negative literals. The $\PSPACE$-complete problem that will be solved by a uniform family of P~systems is the \emph{Quantified SAT} problem in its $3$-CNF variant~\cite{Papadimitriou1993a}, where the output is the truthiness of a quantified formula in 3-CNF.

To encode the input formula $\varphi$ we employ the following encoding:
\begin{itemize}
  \item There are $8\binom{n}{3}$ bits, one for each possible clause. The $i$-th bit set to one means that the $i$-th clause (in lexicographic order) is present in $\varphi$.
  \item There are then $n$ bits, one for each variable, where the $i$-th bit set to one means that $x_i$ is universally quantified. If the $i$-th bit is set to $0$ then $x_i$ is existentially quantified.
\end{itemize}

The previous encoding allows to know from the length of the input the number of variables present in $\varphi$; a similar encoding was already used in~\cite{Porreca2011b}, which we refer to for additional details. Therefore, the Turing machine $F$ of the uniformity condition is able to construct the P~system with the knowledge of $n$ and not only of the length of $\varphi$.

We are going to define a set of symbols, each one denoting a logarithmic number of quantifiers. We assume, without loss of generality, that the variables appear quantified in the order $x_1, x_2, \ldots, x_n$, so that $Q_i$ quantifies variable $x_i$. If this is not true a simple renaming of the variables can provide a formula satisfying this condition. We also assume that the number of variables $n$ is divisible by $k = \lceil \log_2 n \rceil$; if this does not hold then we apply a padding, that in any case add no more than $2k$ variables universally quantified. Let $\ell = \frac{n}{k}$ and let $\BigQ_j$ for $1 \le j \le \ell$ be the string of quantifiers $Q_{(j-1)k + 1} Q_{(j-1)k + 2} \cdots Q_{jk}$ where each of the $\frac{n}{\log_2 n}$ sequences is made of $k$ quantifiers. For example, for a formula $\exists x_1 \forall x_2 \forall x_3 \exists x_4 \forall x_5 \exists x_6 \varphi^\star$ with $6$ variables, $\lceil \log_2 6 \rceil = 3$, thus $\BigQ_1 = \exists x_1 \forall x_2 \forall x_3$ and $\BigQ_2 = \exists x_4 \forall x_5 \exists x_6 $.

Notice that, for a fixed $j \in \set{1, \ldots, \ell}$, the values that can be assumed by $\BigQ_j$ are only polynomial with respect to $n$: each of the quantifiers $Q_{(j-1)k + 1}, \ldots, Q_{jk}$ can only be universal or existential and the variable it quantifies over are only $k$ in number. Therefore, $\BigQ_j$ can assume at most $2^k < 2n$ distinct values.

\subsection{Initial Configuration of the P~System}
\label{sec:initial-conf} 

The initial structure of the P~system constructed by machine $F$ of the uniformity condition consists of $\ell + 1$ linearly nested membranes. The membranes are labelled from $1$ (the outermost membrane) to $\ell+1$ (the innermost membrane). Intuitively, the first $\ell$ non-elementary membranes will be used to evaluate the formula $\varphi$, while the innermost (elementary) membrane will be used to evaluate the formula $\varphi^\star$ for a given assignment.

For each $i \in \set{2, \ldots, \ell + 1}$, the membrane with label $i$ contains objects representing all the variables quantified in $\BigQ_{i-1}$. For example, the membrane with label $2$ contains, in the initial configuration, the objects $x_1, \ldots, x_k$, the membrane with label $3$ contains $x_{k+1}, \ldots, x_{2k}$, and so on. Notice that, since the variables appear quantified in order and the number of variables can be obtained from the length of the input, this construction can be performed by machine $F$ of the uniformity condition. Furthermore, the elementary membrane contains the object $\rset$.

The actual formula $\varphi$ is encoded by two sets of symbols. The quantifiers are grouped in contiguous strings, each one containing $k$ of them and encoded in the objects $\BigQ_1, \ldots, \BigQ_\ell$. For each clause $C_1, \ldots, C_k$ in $\varphi^\star$ the object $C_{i,i}$ is present, obtained by subscripting the clause $C_i$ with $i$, its position inside the formula $\varphi^\star$. For the sake of simplicity we will use the notation $C_i$ in both cases, when no confusion arises. Both kinds of objects are placed in the outermost membrane.  

\subsection{Generation of the Assignments}
\label{sec:assignments}

The generation of the assignments is performed in multiple steps. First of all, each object $\BigQ_j$ representing the $j$-th string of quantifiers must be sent into membrane $j$, whereas the objects $C_1, \ldots, C_m$ representing the clauses of $\varphi^\star$ must be sent into the innermost membrane. Finally, the objects $x_1, \ldots, x_n$ trigger the necessary membrane divisions that allow us to obtain a full $2^k$-ary tree of depth $\ell$. As a result there will still be $2^n$ elementary membranes, one for each possible assignment.

At the beginning of the computation the objects $\BigQ_2, \ldots, \BigQ_\ell$ are all sent in by rules of type~\ref{eq:send-in-quantifiers}. Once object $\BigQ_j$ enters the membrane with label $j$, by rules of typee~\ref{eq:write-quantifier} it is rewritten to a junk object, changing the membrane charge to $(\BigQ_j, 0)$, to record which quantifiers will be evaluated in that membrane and at the moment $0$ objects representing clauses have entered the membrane. Object $\BigQ_1$ is treated differently since it is already inside membrane $1$, so it sets the charge by send out instead of being sent in (rule~\ref{eq:send-out-first-quantifier}):
\begin{align}
  \label{eq:send-in-quantifiers}
  & \psendin{i}{(\BigQ_i, 0)}{\BigQ_j}{(\BigQ_i, 0)}{\BigQ_j} & \text{for $1 < j < i \le \ell$}\\
  \label{eq:write-quantifier}
  & \psendin{j}{0}{\BigQ_j}{(\BigQ_j,0)}{\rubbish} & \text{for $1 < j \le \ell$}\\
  \label{eq:send-out-first-quantifier}
  & \psendout{1}{0}{\BigQ_1}{(\BigQ_1,k,0,0)}{\rubbish} &
\end{align}
The charge $(\BigQ_1,k,0,0)$ represents the fact that the quantifiers in $\BigQ_1$ needs to be evaluated, that currently the $k$-th one is the first that will be evaluated; the meaning of the other two values will be clarified in the following, when the evaluation procedure for the quantifiers is described.

Once all objects of type $\BigQ_j$ are in place, the objects $C_1, \ldots, C_m$ are sent in from the outermost to the innermost membrane in order:
\begin{align}
  \label{eq:send-in-clauses}
  & \psendin{j}{(\BigQ_j,q)}{C_i}{(\BigQ_j,q+1)}{C_i}
  & \text{for $1 < j \le \ell$, $0 \le q < m-1$, and $1 \le i \le m$}\\
  \label{eq:send-in-clauses-bottom}
  & \psendin{\ell+1}{q}{C_i}{q+1}{C_i}
  & \text{for $1 < j \le \ell$, $0 \le q < m-1$, and $1 \le i \le m$}\\
  \label{eq:send-in-last-clause}
  & \psendin{j}{(\BigQ_j,m-1)}{C_i}{(\BigQ_j,k,0,0)}{C_i}
  & \text{for $1 < j \le \ell$ and $1 \le i \le m$}\\
  \label{eq:send-in-last-clause-bottom}
  & \psendin{\ell+1}{m-1}{C_i}{(n - k, 0)}{C_i}
  & \text{for $1 \le i \le m$}
\end{align}
Rules of types~\ref{eq:send-in-clauses} and~\ref{eq:send-in-clauses-bottom} send in the objects representing the first $m-1$ clauses. When an object $C_i$ enters a membrane, it changes the charge to record that another clause has entered. The last clause to enter, when being sent into  non-elementary membrane, it  modifies the charge to allow for the next phase of the assignment generation to start (rules of type~\ref{eq:send-in-last-clause}). When it enters the innermost membrane (rules of type~\ref{eq:send-in-last-clause-bottom}) the charge is changed to record the number of objects of type $\true_i$ and $\false_i$ that must still enter the membrane before starting the evaluation of the assignment.

Once the charges of the non-elementary membranes have been set to the form $(\BigQ_j, k, 0, 0)$, the division process can start. Since all membranes except the outermost one already contain $k$ objects representing the variables of $\varphi$, the charge set by the entrance of the object $C_m$ is sufficient to trigger the applicability of the following rules:
\begin{align}
  \label{eq:divide-non-elementary}
  & \pdivide{j}{(\BigQ_j, k, 0, p)}{x_i}{(\BigQ_j, k, 0, p)}{\false_{i,k}}{(\BigQ_j, k, 0, p + h(i,j))}{\true_{i,k}}
  & \text{for $1 \le j \le \ell$ and $1 \le i \le n$} \\
  \label{eq:divide-elementary}
  & \pdivide{\ell+1}{(n - k, p)}{x_i}{(n - k, p)}{\false_i}{(n - k, p + h(i,j))}{\true_i} 
  & \text{for $1 \le i \le n$}
\end{align}
where $h(i,j)$ is defined as $2^{k - (i - (j-1)k)}$. That is, $p$ can be seen as a $k$-bit number, ranging from $0$ to $2^k-1$, where $h(i,j)$ sets to $1$ the bit corresponding to the position of the variable $x_i$ in the current string $\BigQ_j$ of quantifiers.
For non-elementary membranes, rules of type~\ref{eq:divide-non-elementary} perform the division and update the charge to allow each membrane resulting from the division to have a unique identifier $p$ that can be used to order all the membranes with the same label $j$ resulting from division that are located inside the same membrane. The objects $\true_{i,k}$ and $\false_{i,k}$ will be rewritten $k+1$ times before entering into the inner membranes. This allows for the division phase to be completely performed before the send-in happens. For elementary membranes, rules of type~\ref{eq:divide-elementary} perform the division, also adding the identifier $p$. The result of this division will be a complete tree with a branching factor of $2^k$, due to the $k$ division that happened inside each  membrane.

Finally, it is necessary for the generated assignments to move into the elementary membranes to be evaluated. Since each non-elementary membrane has $2^k$ children, it is necessary to generate enough copies of each assignment to enter all the children membranes; this is made through rules of type~\ref{eq:duplicate}. To avoid any conflict with the division process, each object representing the assignment waits for $k$ steps before being duplicated using rules of type~\ref{eq:counts-to-zero}:
\begin{align}
  \label{eq:counts-to-zero}
  & \pevolve{j}{(\BigQ_j, k, 0, p)}{\alpha_{i.t}}{\alpha_{i,t-1}} &
  \text{for $1 \le t \le k$, $1 \le i \le n$, $1 \le j \le \ell$, and $\alpha \in \set{\true, \false}$} \\
  \label{eq:duplicate}
  & \pevolve{j}{(\BigQ_j, k, 0, p)}{\alpha_{i,0}}{\underbrace{\alpha_i\; \cdots\; \alpha_i}_{\text{$2^k$ times}}} &
  \text{for $1 \le j \le \ell$, $1 \le i \le n$, and $\alpha \in \set{\true,\false}$}
\end{align}
Once the duplication process has been performed, the objects can actually be sent in, either in a non-elementary membrane (rules of type~\ref{eq:send-in-assignment}), where the rewriting process of rules of types~\ref{eq:counts-to-zero} and~\ref{eq:duplicate} will be repeated, or inside an elementary membrane (rules of type~\ref{eq:send-in-assignment-in-elementary}), where the counter for the missing variable assignments present in the charge will be decremented by one:
\begin{align}
  \label{eq:send-in-assignment}
  & \psendin{j}{(\BigQ_j, k, 0, p)}{\alpha_i}{(\BigQ_j, k, 0, p)}{\alpha_{i,k}}
  & \text{for $1 \le i \le k$, $1 \le j \le \ell$, and $\alpha \in \set{\true, \false}$} \\
  \label{eq:send-in-assignment-in-elementary}
  & \psendin{\ell+1}{(n-i, p)}{\alpha_i}{(n-i-1, p)}{\alpha_i}
  & \text{for $1 \le i \le n$ and $\alpha \in \set{\true, \false}$}
\end{align}
Notice that since the number of objects that can be involved in send-in rules present in each membrane is $2^k$, which is also the number of membranes where they can enter, they distribute uniformly (i.e., one per membrane) across all children membranes. At the end of the process, once all assignments are inside the $2^n$ elementary membranes, the charge of all the elementary membranes will be of the form $(0,p)$. The appearance of this charge will make it possible to start the evaluation of the assignments in the next phase.

Notice how the entire process of generating the assignments requires only polynomial time. Each membrane performs at most $k$ divisions (requiring $k$ time steps) and each membrane receives (via send-in) only a polynomial amount of objects (bounded above by the sum of the number of variables, the number of clauses, and the number of quantifiers). Furthermore, the number of possible types of objects and charges remains polynomial during the entire process.

\subsection{Evaluation of Assignments}
\label{sec:assignment-evaluation}

The assignment evaluation is performed inside the elementary membranes. The main idea is that the objects representing the assignment exit one at a time, writing their value in the charge of the membrane. The objects representing the clauses of $\varphi^\star$ are rewritten according to the different charges. If all the clauses are satisfied once the last object of the assignment has been sent out, then that assignment satisfies $\varphi^\star$. Otherwise, the assignment does not satisfy $\varphi^\star$.

For all $p \in \set{0, \ldots, 2^k - 1}$ the following rules are present in the system: 
\begin{align}
  \label{eq:begin-send-out-assignment}
  & \psendout{\ell+1}{(0,p)}{\alpha_1}{(\alpha_1,p)}{\rubbish} & \text{for $\alpha \in \set{\true, \false}$}\\
  \label{eq:send-out-assignment}
  & \psendout{\ell+1}{(\alpha_{i-1},p)}{\beta_i}{(\beta_i,p)}{\rubbish} & \text{for $\alpha, \beta \in \set{\true, \false}$}
\end{align}
These rules send-out as junk all the objects representing the assignment of the formula $\varphi^\star$, starting from the first one (rules of type~\ref{eq:begin-send-out-assignment}), an then sending out all of them one at a time and in order (rules of type~\ref{eq:send-out-assignment}). This allows the following rules to ``read'' the assignment from the charge and rewrite an object representing a clause $C_j$ if one of the variables has an assignment that satisfies it:
\begin{align}
  \label{eq:clause-positive}
  & \pevolve{\ell+1}{(\true_i,p)}{C_j}{\yes_p'} & \text{if $x_i \in \pos(C_j)$}\\
  \label{eq:clause-negative}
  & \pevolve{\ell+1}{(\false_i,p)}{C_j}{\yes_p'} & \text{if $x_i \in \npos(C_j)$}
\end{align}

Once all the objects representing the assignment have been sent out, the object $\rset$ is sent out to change the charge of the membrane using rules of type~\ref{eq:eval-end}. Then the objects $\yes'_p$, representing the clauses that have been satisfied, wait one steps by rewriting themselves in $\yes_p$ using rules of type~\ref{eq:wait-one-step}. At the same time, if at least one object representing one of the clauses has not been rewritten, it exits as $\no_{k,p}$, signalling that the assignment did not satisfy the formula $\varphi^\star$, and sets the charge to $\rubbish$, inhibiting the application of any other rule (rules of type~\ref{eq:clause-unsatisfied}). If this does not happen, then one of the $\yes_p$ objects is sent out as $\yes_{k,p}$, signalling that the assignment did satisfy $\varphi^\star$ (rules of type~\ref{eq:assignment-satisfy}):
\begin{align}
  \label{eq:eval-end}
  & \psendout{\ell+1}{(\alpha_m,p)}{\rset}{\rset}{\rubbish}
  & \text{for $\alpha \in \set{\true, \false}$}\\
  \label{eq:wait-one-step}
  & \pevolve{\ell+1}{\rset}{\yes_p'}{\yes_p} & \\
  \label{eq:clause-unsatisfied}
  & \psendout{\ell+1}{\rset}{C_j}{\rubbish}{\no_{k,p}} & \text{for $1 \le j \le m$}\\
  \label{eq:assignment-satisfy}
  & \psendout{\ell+1}{\rset}{\yes_p}{\rubbish}{\yes_{k,p}} &
\end{align}

The evaluation of an assignment according to these rules requires a time that is linear with respect to the number of variables $n$. Furthermore, since all the involved quantities (e.g., possible values for $p$, number of clauses with three literals) are polynomial with respect to $n$, the number of rules to be defined is also polynomial. Therefore, the evaluation of the assignments requires a polynomial time and all rules necessary to perform it can be constructed in polynomial time. 

\subsection{Quantifiers}
\label{sec:quantifiers}

For the description of the following rules it is useful to introduce the concept of \emph{quantification tree}, that is, the ``shape'' of the computation performed to actually decide if $\varphi$ is a valid formula. This is a complete binary tree of depth $n$ where to each internal level is associated a quantifier of $\varphi$. In particular $Q_1$ is associated to the root (depth $0$), $Q_2$ is associated to the nodes at depth $1$, and so on until the leaves are reached. The leaves are labeled with either $\true$ or $\false$; the exact value depends on the satisfiability of $\varphi^\star$ given an assignment obtained by looking at the path from the root to the particular leaf of the tree ($x_i$ is $\false$ if to reach depth $i$ from depth $i-1$ the path went to the left and $\true$ if it went to the right). To establish the validity of $\varphi$ each internal node acts either as an $\land$ gate (if the quantifier associated to that depth is a universal one) or as an $\lor$ gate (if the associated quantifier is existential). In the following, we will say that the evaluation of the quantifiers moves up a level or moves to other siblings on the same level. This is to be interpreted as a movement on this quantification tree, which is the general method employed to establish the validity of $\varphi$ by this algorithm. However, while usually this entire quantification tree is explicitly represented by the membrane structure, here it is partially present in the membrane structure and partially ``simulated'' by the sequential application of rules inside the non-elementary membranes. Since each level of the membrane structure of this construction ``compresses'' $\log_2 n$ levels of a traditional construction, the resulting tree has a depth that is reduced by a factor of $\frac{1}{\log_2 n}$, thus obtaining a depth of $O\left(\frac{n}{\log_2 n}\right)$.

The evaluation of the quantifiers is performed by alternating two steps: one internal to a membrane and one where the results are sent out to the parent membrane. Each non-elementary membrane receives from its $2^k$ children a result, either $\yes$ or $\no$, of a partial evaluation of the formula $\varphi$. These results are numbered from $0$ to $2^k - 1$ and are treated like the leaves of a complete binary tree of depth $k$, each level of the tree representing a different quantifier of $\BigQ_j$; that is, a fragment of the quantification tree. The results are combined two at a time in a sequential manner, according to the quantifiers of $\BigQ_j$. For example the first $2^k$ results are combined with $\land$ or $\lor$ (depending on the last quantifier of $\BigQ_j$) to obtain $2^{k-1}$ results that will be further combined until a single one is produced. This result of the evaluation of all the quantifiers in $\BigQ_j$ is then sent out to the parent membrane. In the case of the outermost membrane, this is actually the result of evaluating the entire formula $\varphi$. While more involved, this procedure is similar to the one usually employed for solving \QSAT with P~systems with a linear depth with respect to $n$. The main difference is that, here, instead of exploiting a deeper membrane structure, it is necessary to ``compress'' some levels of the membrane structure and perform part of the evaluation sequentially inside a single membrane instead of using multiple nested membranes.

In the following, we assume that $0 \le c < 2^r-1$ and $0 \le p < 2^k$. Every membrane with label $j \in \set{1, \ldots, \ell}$ has associated the following type of rules:
\begin{align}
  \label{eq:send-out-eval}
  & \psendout{j}{(\BigQ_j, r, c, p)}{\alpha_{r, c}}{(\BigQ_j, r, c+1, p, \alpha)}{\rubbish} 
  &\text{for $0 \le r \le k$ and  $\alpha \in \set{\yes, \no}$}
\end{align}
Here the charge contains two indices, $r$ and $c$, that indicate, respectively, that the current evaluation is of the $r$-th quantifier of $\BigQ_j$ and that the $c$-th result is the one that will be read. The values $r$ and $c$ can be interpreted as coordinates in (a fragment of) the quantification tree, where $r$ denotes the depth and $c$ a node among the ones at depth $r$. To complete the evaluation, the $(c+1)$-th result should be combined with the $c$-th one, as follows:
\begin{align}
  \label{eq:eval-rewrite}
  & \pevolve{j}{(\BigQ_j, r, c, p, \alpha)}{\beta_{r, c}}{\gamma_{r-1,\lfloor \frac{c}{2} \rfloor}, \reset}
  & \text{for $1 \le j \le \ell$ and $\alpha, \beta \in \set{\yes, \no}$}
\end{align}
where $\gamma$ is $\alpha \land \beta$ if the $r$-th quantifier in $\BigQ_j$ is a universal one and $\alpha \lor \beta$ if it is an existential one. Where $alpha$ and $\beta$ are two siblings in the evaluation tree and are combined to obtain the value $\gamma$ of their parent. The object $\reset$ appearing in rules of type~\ref{eq:eval-rewrite} is used to signal (by being sent out) that the object representing $\alpha \land \beta$ (for a universal quantifier) or $\alpha \lor \beta$ (for an existential quantifier) has been produced and, thus, that the evaluation can continue. This action is performed by the following rules:
\begin{align}
  \label{eq:move-eval-right}
  & \psendout{j}{(\BigQ_j, r, c, p, \alpha)}{\reset}{(\BigQ_j, r, c+1, p)}{\rubbish}
  & \text{for $1 \le j \le \ell$ and $\alpha \in \set{\yes, \no}$} \\
  \label{eq:move-eval-up}
  & \psendout{j}{(\BigQ_j, r, 2^r-1, p, \alpha)}{\reset}{(\BigQ_j, r-1, 0, p)}{\rubbish}
  & \text{for $1 \le j \le \ell$ and $\alpha \in \set{\yes, \no}$} \\
  \label{eq:move-eval-out}
  & \psendout{j}{(\BigQ_j, 1, 1, p)}{\alpha_{0,0}}{\rubbish}{\alpha_{k, p'}}
  & \text{for $1 < j \le \ell$ and $\alpha \in \set{\yes, \no}$} \\
  \label{eq:produce-result}
  & \psendout{1}{(\BigQ_j, 1, 1, p)}{\alpha_{0,0}}{\rubbish}{\alpha}
  & \text{for $\alpha \in \set{\yes, \no}$}
\end{align}
Rules of type~\ref{eq:move-eval-right} modify the charge to evaluate the next two results in the quantification tree while remaining at the same level. Once a level has been exhausted, the evaluation moves up by means of rules or type~\ref{eq:move-eval-up}. Once the entire part of the quantification tree that is ``simulated'' inside a membrane of label $j$ is exhausted, rules of type~\ref{eq:move-eval-out} move the result to the parent membrane. The result will be at the bottom ($k$-th) level of the quantification tree of the parent membrane and its position among all the other results will be given by value $p$ that was previously stored in the membrane charge. Finally, if the membrane where the evaluation ended is the outermost one, then the produced result is actually the result of the entire computation, and it is sent out as either $\yes$ or $\no$.

Notice that the part of the quantification tree that is ``simulated'' by each non-elementary membrane is polynomial in the number of nodes that it contains. Therefore, the sequential evaluation performed in this construction still requires only polynomial time. Since among membranes at the same level all evaluations are performed in parallel, the time needed to produce the result of the entire computation multiplies this time by a factor that depends only on the depth of the membrane structure. Therefore, the P~system resulting from this construction is able to produce an answer in a time which is polynomial with respect to the input size.

\subsection{Main Result}

The construction presented here shows that \QSAT in its $3$-CNF variant can be solved with a nesting depth that is sublinear with respect to the number of variables. This is the first solution, as far as the authors know, that goes below a linear nesting depth. In particular:

\begin{theorem}
  Uniform families of P~systems with active membranes with charges and weak non-elementary division rules can solve the \QSAT problem in $3$-CNF form, where the quantified formula given as input has $n$ variables, using a depth of $O(\frac{n}{\log n})$ \qed
\end{theorem}

Even if the construction employed uses polynomial charges, instead of the usual $3$, it has already been proved that one system can be converted into the other with only a polynomial slowdown and no increase in depth~\cite{Leporati2017b}. We want to remark that the availability of additional charges allowed a more compact and easier construction.

\section{Conclusions}
\label{sec:conclusions}

While solving $\PSPACE$-complete problems, \QSAT in particular, with P~systems with active membranes with charges employing weak non-elementary division rules is not a new result, the construction provided here is the first one where the nesting depth of the membrane structure is sub-linear with respect to the number of variables in input. This is a first step in the direction of characterising the power of families of P~systems with sublinear depth. While we have provided a construction reducing the depth needed in a specific problem (\QSAT with a formula in $3$-CNF), but it is still open what is the impact of this result in term of complexity classes. We want to remark that this construction cannot be directly employed to reduce the depth of the membrane structure below $O\left(\frac{n}{\log_2 n}\right)$, because we are already employing a polynomial number of charges, types of objects, and rules. Employing the same method to further reduce depth, for example to $O\left(\frac{n}{(\log_2 n)^2}\right)$, would require a superpolynomial number of charges, object types, and rules -- thus violating the uniformity condition of the family. 

The remaining investigation work is vast: a complete characterisation of the families of constant depth is still in the work and other classes, like the one of families of logarithmic depth, are unexplored. Charting this unknown space of complexity classes is a long-term objective that will probably be necessary to attain in order to completely understand the complex interaction between nesting depth and computational power in P~systems.

\bibliographystyle{splncs03}
\bibliography{Bibliography}

\end{document}